\documentclass[conference]{IEEEtran}

\usepackage[T1]{fontenc}
\usepackage{amsmath}

\usepackage{txfonts}

\newcommand{\qed}{\hspace*{\fill}\rule{1ex}{1ex}}
\newcommand{\rank}{\mathrm{rank}}

\title{Secure Multiplex Network Coding}
\author{\IEEEauthorblockN{Ryutaroh Matsumoto}\IEEEauthorblockA{%
Department of Communications and Integrated Systems,\\
Tokyo Instiutte of Technology, 152-8550 Japan}
 \and \IEEEauthorblockN{Masahito Hayashi}\IEEEauthorblockA{%
Graduate School of Information Sciences,\\
Tohoku University, 980-8579 Japan\\
and
Centre for Quantum Technologies,\\
 National University of Singapore,\\
 3 Science Drive 2, Singapore 117542}}
\date{\today}

\newtheorem{definition}{Definition}

\newtheorem{theorem}[definition]{Theorem}

\newtheorem{remark}[definition]{Remark}
\newtheorem{lemma}[definition]{Lemma}

\allowdisplaybreaks

\begin{document}
\maketitle
\thispagestyle{plain}
\begin{abstract}
In the secure network coding for multicasting,
there is loss of information rate due to inclusion of random bits at the source node.
We  show a method to eliminate that loss of information rate
by using multiple statistically independent messages to be kept secret from an eavesdropper.
The proposed scheme is an adaptation of Yamamoto et~al.'s secure multiplex coding \cite{yamamoto05}
to the secure network coding \cite{cai02b,caiyeung11,silva08}.
\end{abstract}



\section{Introduction}\label{sec1}
Network coding \cite{ahlswede00} attracts much attention recently
because it can offer improvements in several metrics, such as throughput
and energy consumption, see \cite{fragouli07b,fragouli07a}.
On the other hand, the information theoretic security \cite{liang09} also
attracts much attention  because it offers security that does not depend on
a conjectured difficulty of some computational problem.

A juncture of the network coding and the information theoretic security
is the secure network coding \cite{cai02b,caiyeung11}, which prevents an eavesdropper, called Eve,
from knowing the message from the legitimate sender,
called Alice, to the legitimate receivers
by eavesdropping intermediate links up to a specified number in a network.
It can be seen \cite{rouayheb07} as a network coding counterpart of the traditional wiretap channel coding
problem considered by Wyner \cite{wyner75} and subsequently others \cite{liang09}.
In both secure network coding and coding for wiretap channels,
the secrecy is realized by including random bits statistically
independent of the secret message into the transmitted signal by Alice
so that the secret message
becomes ambiguous to Eve. The inclusion of random bits, of course,
decreases the information rate.
In order to get rid of the decrease in the information rate,
Yamamoto et~al.\ \cite{yamamoto05}
proposed the secure multiplex coding for wiretap channels, in which
there is no loss of information rate.
The idea of Yamamoto et~al.\  is as follows:
Suppose that Alice has $T$ statistically independent messages
$S_1$, \ldots, $S_T$.
Then $S_1$, \ldots, $S_{i-1}$, $S_{i+1}$, \ldots,
$S_T$ serve as the random bits making $S_i$ ambiguous to Eve, for each $i$.
The purpose of this paper is to do the same thing with the secure
network coding as Yamamoto et~al.\ \cite{yamamoto05}.

Independently and simultaneously,
Bhattad and Narayanan \cite{bhattad05} proposed the
weakly secure network coding, whose goal is also to get rid of
the loss of information rate in the secure network coding.
Their method \cite{bhattad05} ensures that the mutual information
between $S_i$ and Eve's information is zero for each $i$.
As drawbacks,
the construction depends on the network topology and coding at intermediate
nodes, and the computational complexity of code construction is large.
In order to remove these drawbacks,
Silva and Kschischang \cite{silva09} proposed the universal
weakly secure network coding, in which they showed an efficient
code construction that can support up to two $\mathbf{F}_q$-symbols in each $S_i$ and
is independent of the network topology and coding at intermediate
nodes. They \cite{silva09} also  showed the existence of universal
weakly secure network coding with more than two $\mathbf{F}_q$-symbols
in $S_i$, but have not shown an explicit construction.

We shall propose a construction of
secure multiplex network coding
that is an adaptation of Yamamoto et~al.'s idea \cite{yamamoto05}
to the network coding.
However, we relax an aspect of the security requirements
traditionally used in the secure network coding.
In previous proposals of secure network coding \cite{cai02b,bhattad05,harada08,silva09,silva08}
it is required that the mutual information to the eavesdropper is exactly
zero.
We relax this requirement by regarding sufficiently small mutual information
to be acceptable.
This relaxation is similar to requiring the bit error rate
to be sufficiently small instead of strictly zero.
Also observe that our relaxed criterion is much stronger than
one commonly used in the information theoretic security \cite{liang09}.
Our construction can realize arbitrary small mutual information
if coding over sufficiently many time slots is allowed.

There are several reasonable models of the eavesdropper Eve.
In the traditional model used in \cite{cai02b,caiyeung11,harada08,silva09,silva08},
Eve can arbitrary choose the set of $\mu$ eavesdropped links after
learning the structure of network coding and the set of eavesdropped links
is assumed to be constant during transmission of one coding block.
The secure network coding is required to leak no information with
every set of $\mu$ eavesdropped links.
We call this model as the traditional eavesdropping model.
We shall show that the proposed scheme is universal secure in
Section \ref{sec32} in the sense of \cite{silva09,silva08}
under the traditional eavesdropping model.

However, it is observed in \cite{shioji10} that there is difficulty in implementation over the current Internet architecture
to keep the set of eavesdropped links constant even when
the set of eavesdropped links is physically constant.
Thus, we consider another model of the eavesdropper in which
the set of eavesdropped links is statistically distributed
independent of the structure of transmitter,
the number of eavesdropped links is $\mu$ per unit time, and
the set of eavesdropped links is allowed to be time-varying.
We call this second model as the statistical eavesdropping model.
We shall also show that the mutual information is small
averaged over \emph{any probability distribution} of network coding statistically independent of
any other random variables,
instead of the mutual information being
small with every network coding as done in \cite{cai02b,caiyeung11,harada08,silva09,silva08}.
Since the network coding is often constructed in the
random manner \cite{ho06},
considering the probability distribution of network coding
and requiring the averaged mutual information being small
make sense.
We also define a Shannon theoretic capacity region
of secure multiplex network coding
and show that the proposed construction can achieve that capacity region.

Although Harada and Yamamoto \cite{harada08} have not explicitly stated,
the adaptation of the secure multiplex coding \cite{yamamoto05} to the
secure network coding \cite{cai02b,caiyeung11} can be done by their
strongly secure network coding \cite{harada08}. The difference between
our proposed scheme and the previous works \cite{bhattad05,harada08,silva09} is as follows:
\begin{itemize}
\item The computational complexity of constructing network coding
is huge in \cite{bhattad05,harada08}, while the computational complexity of
code construction in the
proposed scheme is that of selecting a random nonsingular linear matrix.
\item The construction of network coding in \cite{bhattad05,harada08} depends
on the underlying network topology, while the proposed scheme is independent
of it and universal secure in the sense of \cite{silva09,silva08} under the
traditional eavesdropping model (see Section \ref{sec32}).
\item The explicit construction of universal weakly secure network coding
in \cite{silva09} supports up to two $\mathbf{F}_q$-symbols in each
secret message $S_i$ and guarantees that the mutual information between each $S_i$
and Eve's information is zero, while the proposed scheme has no limitation
on the size of $S_i$ and can make the mutual information between 
any collection $(S_i : i \in \mathcal{I})$ and Eve's information arbitrary
small\footnote{The mutual information turned out to be exactly
zero, see Appendix \ref{app:b}.} provided that the total information rate of $(S_i : i \in \mathcal{I})$
is not too large relative to $\mu$.
\item When the total information rate of $(S_i : i \in \mathcal{I})$ is large
relative to $\mu$, the mutual information to Eve becomes positive,
but \cite{bhattad05,silva09} do not evaluate how large it is
nor their smallest possible value.
The proposed scheme realizes asymptotically
the smallest possible value of mutual
information with every collection $(S_i : i \in \mathcal{I})$
\emph{simultaneously,} as well as \cite{harada08}.
\end{itemize}

This paper is organized as follows:
Section \ref{sec2} reviews related results used in this paper.
Section \ref{sec3} introduces the strengthened version of the privacy
amplification theorem and the proposed scheme
for secure network coding, then proves the
asymptotic optimality of the latter.
Section \ref{sec4} concludes the paper.

\section{Preliminary}\label{sec2}
\subsection{Model of network coding}
As in \cite{bhattad05,cai02b,caiyeung11,harada08,silva09,silva08}
we consider the single source multicast.
The network model is either acyclic or cyclic, and
each link has either no delay or unit delay.
We assume the linear network coding \cite{li03},
and a link carries single $\mathbf{F}_q$ symbol per time slot.
The linear combination coefficients at each node are fixed so that
every legitimate receiver can receive $n$ symbols per time slot from the source.
Linear combination coefficients at a node are allowed to
change at each time slot except in Section \ref{sec32}.
We shall only consider the eavesdropper Eve and
forget about the legitimate receivers.
We shall propose a coding method encoding information over $m$ time slots at
the source node. Therefore, the source node transmit $(m \times n)$ $\mathbf{F}_q$
symbols in a single coding block.
Eve can eavesdrop $\mu$ links per time slot.
We assume $\mu \leq n$ throughout this paper.
The total number of eavesdropped links is therefore $m \mu$.

\subsection{Two-universal hash functions}
We shall use a family of two-universal hash functions \cite{carter79}
for the privacy amplification theorem introduced later.

\begin{definition}\label{def:twouniv}
Let $\mathcal{F}$ be a set of functions from a finite set $\mathcal{S}_1$ to
another finite  set $\mathcal{S}_2$,
and $F$ the uniform random variable on $\mathcal{F}$. If for any $x_1 \neq x_2
\in \mathcal{S}_1$ we have
\[
\mathrm{Pr}[F(x_1)=F(x_2)] \leq \frac{1}{|\mathcal{S}_2|},
\]
then $\mathcal{F}$ is said to be a \emph{family of two-universal hash functions}.
\end{definition}


\section{Construction of secure multiplex network coding}\label{sec3}
\subsection{Strengthened privacy amplification theorem}
In order to evaluate the mutual information to Eve when the
rate of secret information is large,
we need to strengthen the privacy amplification theorem
originally appeared in \cite{bennett95privacy,hayashi11}
as follows. The original version of the privacy amplification theorems
\cite{bennett95privacy,hayashi11} cannot deduce Eq.\ (\ref{eq:ub7}) while
Theorem  \ref{thm2} can.
\begin{theorem}\label{thm2}
Let $X$ and $Z$ be discrete random variables on finite sets $\mathcal{X}$ and
$\mathcal{Z}$, respectively, and $\mathcal{F}$ be a family of two-universal
hash functions from $\mathcal{X}$ to $\mathcal{S}$.
Let $F$ be the uniform random variable on $\mathcal{F}$ statistically independent of
$X$ and $Z$. Then we have
\begin{equation}
\mathbf{E}_f \exp(\rho I(F(X);Z|F=f))  \leq   
1+ |\mathcal{S}|^\rho\mathbf{E}[P_{X|Z}(X|Z)^\rho]\label{hpa1}
\end{equation}
for all $0 \leq \rho \leq 1$,
where $I$ denotes the (conditional) mutual information
as defined in \cite{cover06}.
We use the natural logarithm for
all the logarithms in this paper,
which include ones implicitly appearing in entropy and mutual information.
Otherwise we have to adjust the above inequality.
Proof will be given in Appendix \ref{app:a}.
\end{theorem}

\subsection{Description of the proposed scheme}\label{sec31}
We assume that we have $T$ statistically independent and
uniformly distributed secret messages, and
that the $i$-th secret message is given as a random variable $S_i$
whose realization is a column vector in $\mathbf{F}_q^{k_i}$.
The sizes $k_i$ are determined later.
We shall also use a supplementary random message
$S_{T+1}$ taking values in $\mathbf{F}_q^{k_{T+1}}$
when the randomness in the encoder is insufficient to make
$S_i$ secret from Eve.
We assume $mn= k_1 + \cdots+k_{T+1}$.
Let $\mathcal{L}$ be the set of all bijective $\mathbf{F}_q$-linear
maps from $\prod_{i=1}^{T+1} \mathbf{F}_q^{k_i}$ to itself,
and $\alpha_{\mathcal{I}}$ be the projection from
$\prod_{i=1}^{T+1} \mathbf{F}_q^{k_i}$ to $\prod_{i\in\mathcal{I}} \mathbf{F}_q^{k_i}$
for $\emptyset \neq \mathcal{I} \subseteq \{1$, \ldots, $T\}$.
In \cite{matsumotohayashi2011isit} we have shown that
the family $\{ \alpha_{\mathcal{I}}\circ L \mid L \in \mathcal{L}\}$
is that of two-universal hash functions for 
all $\emptyset \neq \mathcal{I} \subseteq \{1$, \ldots, $T\}$.
Let $L$ be the uniform random variable on $\mathcal{L}$
statistically independent of $S_1$, \ldots, $S_{T+1}$,
and arbitrary fix nonempty $\mathcal{I} \subseteq \{1$, \ldots, $T\}$.
Define $X$ to be the random variable $L^{-1}(S_1$, \ldots,
$S_T$, $S_{T+1})$.
The source node sends $X$ to its $n$ outgoing links over $m$ time slots.
Our construction just attaches the inverse of a bijective
linear function to an
existing network coding.

By the assumption on Eve, her information can be expressed as
$BX$ by using a $\mu m \times mn$ matrix $B$ over $\mathbf{F}_q$.
We regard $B$ as a random variable (matrix) and its probability
distribution is denoted by $P_B$.
We make no assumption on $P_B$ except that
the rank of $B$ is at most $\mu m$ and that
$B$ is independent of $S_1$, \ldots, $S_{T+1}$, $L$ and $X$,
which means that Eve does not change $B$ by watching the realization of
$L$.
When the random network coding is used and Eve can choose locations of up to $\mu$
eavesdropping links but cannot choose the linear coefficients of random network coding,
the statistical independence assumption on $B$ looks reasonable.
 From
the uniformity assumption on $S_i$,
the conditional distribution $P_{X|L}$ is uniform with every realization of $L$.
This means that $X$ and $L$ are statistically independent and $P_X$ is uniform.
For fixed $z$, the set $\{ x \in \mathbf{F}_q^{mn} \mid
Bx = z \}$ has $q^{mn-\rank B}$ vectors.
Thus, the conditional distribution
$P_{X|BX,L,B}(x|z,\ell,b)$ is the uniform distribution on the set of $q^{mn-\rank b}$
elements with every triple of $z$, $\ell$ and $b$. We have
\begin{equation}
P_{X|BX,L,B}(x|z,l,b) = P_{X|BX,B}(x|z,b) = q^{-(mn-\rank b)}.\label{eq2}
\end{equation}

Arbitrary fix nonempty $\mathcal{I} \subseteq \{1$, \ldots, $T\}$,
denote the collection of random variables $(S_i : i \in \mathcal{I})$
by $S_{\mathcal{I}}$, also fix a realization $b$ of $B$, and let
$k_{\mathcal{I}} = \sum_{i\in\mathcal{I}} k_i$.
Under these notations, we can upper bound the mutual information
$I(S_{\mathcal{I}}; bX|L)$ as
\begin{align}
& \mathbf{E}_\ell \exp(\rho I(S_{\mathcal{I}};BX|B=b,L=\ell))\label{eq:ub1}\\
& \leq 
 1+q^{\rho k_{\mathcal{I}}} \mathbf{E}[P_{X|bX}(X|bX)^\rho]
\textrm{ (by Theorem \ref{thm2})}\nonumber\\
&= 1+q^{\rho k_{\mathcal{I}}} \mathbf{E}[(q^{-(mn-\rank b)})^\rho]
\textrm{ (by Eq.\ (\ref{eq2}))}\nonumber\\
&\leq  1+ q^{\rho k_{\mathcal{I}}} q^{-m\rho (n-\mu)}\textrm{ (because $\rank b
\leq \mu m$)}\nonumber\\
&= 1+q^{-m\rho(n-\mu - k_{\mathcal{I}}/m)} \label{eq:ub2}
\end{align}
for $0 \leq \rho \leq 1$.
One can also see that
\begin{align}
& \mathbf{E}_\ell \rho I(S_{\mathcal{I}};BX|B=b,L=\ell)\label{eq:ub3}\\
& \log \exp (\mathbf{E}_\ell \rho I(S_{\mathcal{I}};BX|B=b,L=\ell))\nonumber\\
&\leq \log \mathbf{E}_\ell \exp ( \rho I(S_{\mathcal{I}};BX|B=b,L=\ell))\nonumber\\
&\leq \log (1+q^{-m\rho(n-\mu - k_{\mathcal{I}}/m)}) \nonumber\\
&\leq q^{-m\rho(n-\mu - k_{\mathcal{I}}/m)}\label{eq:ub4}
\end{align}
for $0 \leq \rho \leq 1$.
Averaging Eqs.\ (\ref{eq:ub1})--(\ref{eq:ub4}) over $b$ we have
\begin{align}
\mathbf{E}_{b,\ell} \rho I(S_{\mathcal{I}};BX|B=b,L=\ell) &\leq q^{-m\rho(n-\mu - k_{\mathcal{I}}/m)}\label{eq:ub101},\\
\mathbf{E}_{b,\ell} \exp(\rho I(S_{\mathcal{I}};BX|B=b,L=\ell)) &\leq 1+q^{-m\rho(n-\mu - k_{\mathcal{I}}/m)}. \label{eq:ub102}
\end{align}

\newpage 
Fix $C_1 > 2 \times (2^T-1)$.
Equation (\ref{eq:ub101}) and the Markov inequality yield that
\begin{align*}
\mathrm{Pr} \mathcal{L}_{\mathcal{I},1} <  1/C_1 
\end{align*}
for any single nonempty $\mathcal{I} \subseteq \{1$, \ldots, $T\}$,
where $\mathcal{L}_{\mathcal{I},1}$ $:=$
$\{ \ell \mid \mathbf{E}_{b} I(S_{\mathcal{I}};BX|B=b,L=\ell) > C_1 \mathbf{E}_{b,\ell} I(S_{\mathcal{I}};BX|B=b,L=\ell) \}$.
Thus,
\[
\mathrm{Pr} \cup_{\mathcal{I}: \mathcal{I}\neq \emptyset} \mathcal{L}_{\mathcal{I},1}
 <   (2^T-1)/C_1.
\]
This means that a realization $\ell$ of $L$
satisfies 
\begin{align}
& \mathbf{E}_b I(S_{\mathcal{I}};BX|B=b,L=\ell) \nonumber \\
& \leq 
C_1 \mathbf{E}_{b,\ell} I(S_{\mathcal{I}};BX|B=b,L=\ell) \nonumber \\
& \leq 
C_1 q^{-m\rho(n-\mu - k_{\mathcal{I}}/m)}/\rho\label{eq:ub5}
\end{align}
for all the $(2^T-1)$ nonempty subsets $\mathcal{I}$ of $\{1$, \ldots, $T\}$
with probability  at least $1-(2^T-1)/C_1$.
Defining another subset $\mathcal{L}_{\mathcal{I},2}$ $:=$
$\{ \ell \mid 
\mathbf{E}_b \exp(\rho I(S_{\mathcal{I}};BX|B=b,L=\ell))
> 
C_1 
\mathbf{E}_{b,\ell} \exp(\rho I(S_{\mathcal{I}};BX|B=b,L=\ell))$,
by Eq.\ (\ref{eq:ub102}) and the Markov inequality we obtain 
\[
\mathrm{Pr} \cup_{\mathcal{I}: \mathcal{I}\neq \emptyset} 
(\mathcal{L}_{\mathcal{I},1} \cup \mathcal{L}_{\mathcal{I},2})
 <   2 (2^T-1)/C_1.
\]
Therefore, 
a realization $\ell$ of $L$ satisfies both Eq. (\ref{eq:ub5}) and
\begin{align}
\mathbf{E}_b \exp(\rho I(S_{\mathcal{I}};BX|B=b,L=\ell))
&\leq  C_1 (1+q^{-m\rho(n-\mu - k_{\mathcal{I}}/m)}).\label{eq:ub6}
\end{align}
with probability  at least $1-2 \times (2^T-1)/C_1$.

Equation (\ref{eq:ub6}) implies 
\begin{align}
& \mathbf{E}_b\frac{I(S_{\mathcal{I}};BX|B=b,L=\ell)}{m}\nonumber\\
& = \frac{1}{m}\mathbf{E}_b\log \exp I(S_{\mathcal{I}};BX|B=b,L=\ell)\nonumber\\
& \leq \frac{1}{m} \log \mathbf{E}_b \exp I(S_{\mathcal{I}};BX|B=b,L=\ell)\nonumber\\
&\leq  \frac{\log C_1}{m\rho} + \frac{1}{m\rho}\log
(1+q^{-m\rho(n-\mu - k_{\mathcal{I}}/m)}) \textrm{ (by Eq.\ (\ref{eq:ub6}))}\nonumber\\
&\leq
 \frac{1+\log C_1}{m\rho} + 
(k_{\mathcal{I}}/m -(n-\mu))\log q, \label{eq:ub7}
\end{align}
for $k_{\mathcal{I}}/m -(n-\mu) \geq 0$,
where in Eq.\ (\ref{eq:ub7}) we used $\log(1+\exp(x)) \leq 1+x$ for $x\geq 0$.
Note that Eq.\ (\ref{eq:ub5}) is minimized at $\rho=1$, because
its partial derivative with respect to $\rho$ is 
\[
-\frac{C_1 q^{-m \rho (n-\mu-k/m)}}{\rho^2}-\frac{C_1 m (n-\mu-k/m) \log (q) q^{-m \rho (n-\mu-k/m)}}{\rho},
\]
which is negative for all $0 < \rho \leq 1$.
By the same reason Eq.\ (\ref{eq:ub7}) is also minimized at $\rho=1$.

Fix $C_2$ with $C_2 > 2 \times (2^T-1)$.
By the same argument as above, we see that 
for a realization $\ell$ of $L$ satisfying both Eqs.\ (\ref{eq:ub5}) and (\ref{eq:ub7}),
with probability at least 
\begin{equation}
1-2 \times (2^T-1)/C_2, \label{eq:cb}
\end{equation}
a realization $b$ of $B$ makes 
\begin{align}
 I(S_{\mathcal{I}};BX|B=b,L=\ell) & \leq  C_1 C_2 q^{-m\rho(n-\mu - k_{\mathcal{I}}/m)}/\rho,\label{eq:ub8}\\
\frac{I(S_{\mathcal{I}};BX|B=b,L=\ell)}{m}
&\leq 
\frac{1+\log C_2 + \log C_1}{m\rho} \nonumber\\*
&\qquad + 
(k_{\mathcal{I}}/m - (n-\mu))\log q.\label{eq:ub9}
\end{align}

Even when we choose large $C_2$ and $C_1$ and make the probability of
$b$ and $\ell$ satisfying Eqs.\ (\ref{eq:ub8}) and (\ref{eq:ub9}) large,
we can make the upper bound (\ref{eq:ub8})
arbitrary small by increasing $m$.
This observation enables us to use a fixed bijective linear
 function $\ell$ that is agreed
between the legitimate sender and the receivers in advance.
The legitimate receiver can apply the agreed $\ell$ to the received information
in order to restore the original messages $S_i$.

One can easily see that
if $\rho=1$ and  $k_{\mathcal{I}} \leq m(n-\mu - \delta_{\mathcal{I}})$ with $\delta_{\mathcal{I}} >0$,
then the upper bound (\ref{eq:ub8}) on the mutual information exponentially converges to zero
as $m \rightarrow \infty$. This means that we can transmit $(n-\mu)$ $\mathbf{F}_q$-symbols in the $i$-th secret message
per time slot with arbitrary small eavesdropped information when the encoding is allowed to
be done over sufficiently many time slots.
On the other hand, if $m(n-\mu) \leq k_{\mathcal{I}} \leq m(R_{\mathcal{I}} - \delta_{\mathcal{I}})$ with $\delta_{\mathcal{I}} >0$ and 
$R_{\mathcal{I}}>0$, then
by the upper bound (\ref{eq:ub9}) we see the mutual information per symbol
is upper bounded by 
\[
\frac{I(S_{\mathcal{I}};BX|B=b,L=\ell)}{m} < R_{\mathcal{I}} - (n-\mu)
\]
for sufficiently large $m$.
In  Section \ref{sec33} we shall show that the proposed scheme is asymptotically optimal by
capacity consideration.

\begin{remark}
The meanings of $C_1$ and $C_2$ are as follows:
At Eqs.\ (\ref{eq:ub101}) and (\ref{eq:ub102}),
there might not exist a realization $\ell$ of $L$ that
satisfies Eqs.\ (\ref{eq:ub101}) and (\ref{eq:ub102})
for all subsets $\mathcal{I}$ of $\{1$, \ldots, $T\}$ simultaneously.
By sacrificing the tightness of the upper bounds, we
ensure the existence of $\ell$ satisfying Eqs.\ (\ref{eq:ub5})
and (\ref{eq:ub6}).
At Eqs.\ (\ref{eq:ub5}) and (\ref{eq:ub6}),
realizations $b$ of $B$ might satisfy Eqs.\ (\ref{eq:ub5}) and (\ref{eq:ub6})
with unacceptably low probability.
Again by sacrificing the tightness of the upper bounds,
we
ensure that the eavesdropping matrix $B$ satisfies
Eqs.\ (\ref{eq:ub8}) and (\ref{eq:ub9}) with comfortably high
probability.
\end{remark}

\subsection{Security analysis of the proposed scheme under the traditional
eavesdropping model}\label{sec32}
In the preceding study of secure network coding \cite{cai02b,caiyeung11,harada08,silva09,silva08},
it is assumed that 
\begin{itemize}
\item the eavesdropper Eve can choose $\mu$ eavesdropped links per
unit time
after learning the structure of network coding, and
\item the set of eavesdropped links is constant during transmission
of one coding block.
\end{itemize}
In Section \ref{sec1} we called the above assumption as the traditional eavesdropping model.
Under the above assumption, the number of possible sets of
eavesdropped links is constant, say $C_E$, independent of $m$.
If we take the random variable $B$ according to the uniform
distribution on the sets of eavesdropped links and
the probability of some event of $B$ is larger than
$1-1/C_E$ then that probability must be one.
Set $C_2$ in Eq.\ (\ref{eq:cb}) such that $1-2 \times (2^T-1)/C_2 > 1 - 1/C_E$,
then we can see that Eqs.\ (\ref{eq:ub8}) and (\ref{eq:ub9})
hold with every realization of $B$.
In addition to this, the proposed construction of coding
does not depend on the network topology nor coding at intermediate nodes.
In that sense, the proposed scheme is universal secure \cite{silva09,silva08}
except that the proposed scheme can make the mutual information
arbitrary small\footnote{%
The mutual information turned out to be exactly zero and
our scheme is exactly universal secure in the sense of \cite{silva09,silva08},
see Appendix \ref{app:b}.}, while \cite{silva09,silva08} make the mutual information
exactly zero.

\subsection{Capacity consideration of the proposed scheme}\label{sec33}
Firstly, let us define the achievable rate tuple and the capacity region
as in \cite{cover06}.
\begin{definition}\label{def:capacity}
Suppose that we are given a sequence of $\mu m \times mn$ random matrices $B_m$
whose distribution $P_{B_m}$ has no restriction except that the rank of $B_m$ is always $\mu m$
and that $B_m$ is independent of any other random variables.
Let $e_m$ be a stochastic encoder from $\prod_{i=1}^T \mathbf{F}_q^{ m\kappa_{i,m}}$ to
$\mathbf{F}_q^{mn}$,
$d_m$ either stochastic or deterministic decoder from $\mathbf{F}_q^{mn}$ to
$\prod_{i=1}^T \mathbf{F}_q^{ m\kappa_{i,m}}$, and
$S_{i,m}$ the uniform random variable on $\mathbf{F}_q^{ m\kappa_{i,m}}$,
for $i=1$, \ldots, $T$ and $m = 1$, $2$, \ldots.
If
\begin{align*}
\lim_{m\rightarrow \infty} \mathrm{Pr}[(S_{1,m},\ldots, S_{T,m}) \neq d_m(e_m(S_{1,m},\ldots, S_{T,m}))] &= 0,\\
\limsup_{m\rightarrow \infty}  I(S_{\mathcal{I},m}; B_me_m(S_{1,m},\ldots,S_{T,m})|B_m)/m &\\
\mbox{ } \leq \max\left\{0,- (n-\mu) + \sum_{i\in\mathcal{I}} R_i \right\},&\\
\liminf_{m\rightarrow \infty} \kappa_{i,m} &\geq R_i,
\end{align*}
for all nonempty subsets $\mathcal{I} \subseteq \{1$, \ldots, $T\}$,
then the rate tuple $(R_1$, \ldots, $R_T)$ is said to be \emph{achievable} for the secure multiplex network coding 
with $T$ secret messages and up to $\mu$ eavesdropped links per time slots represented by random matrices $B_m$,
where $S_{\mathcal{I},m}$ denotes the collection $(S_{i,m} : i\in\mathcal{I})$ of
secret messages $S_{i,m}$.
We also define the \emph{capacity region of
$\mathbf{F}_q$-linear secure multiplex network coding} as the closure
of such rate tuples $(R_1$, \ldots, $R_T)$ over
all the sequences of encoders and decoders.
\end{definition}

\begin{theorem}
The capacity region $\mathbf{F}_q$-linear secure multiplex network coding
is given by rate tuples $(R_1$, \ldots, $R_T)$ such that
\begin{align*}
0&\leq R_i,\\
\sum_{i=1}^T R_i & \leq  n.
\end{align*}
\end{theorem}

\noindent\emph{Proof.}
The fact that every rate tuple given by the above equation
is achievable is already proved in the previous section
under stronger requirements, namely $\mathrm{Pr}[(S_{1,m}$, \ldots,
$S_{T,m}) \neq d_m(e_m(S_{1,m}$, \ldots,
$S_{T,m}))]=0$ for all $m$,
$\limsup_{m\rightarrow \infty}$
$I(S_{\mathcal{I},m}; B_me_m(S_{1,m}$, \ldots,
$S_{T,m})|B_m)/m \leq \max\{0$, $\sum_{i\in\mathcal{I}} R_i - (n-\mu)\}$, and
$\lim_{m\rightarrow \infty}$
$I(S_{\mathcal{I},m}; B_me_m(S_{1,m}$, \ldots,
$S_{T,m})|B_m) = 0$ for $\mathcal{I}$ with $\sum_{i\in\mathcal{I}} R_i < (n-\mu)$.
We have to show the so-called converse part of the coding theorem.
Fix an arbitrary nonempty subset
$\mathcal{I} \subseteq \{1$, \ldots, $T\}$ and a sequence of random matrices $B_m$,
and suppose that we have a sequence of stochastic encoders as defined in Definition
\ref{def:capacity} with
\begin{equation}
\liminf_{m\rightarrow\infty}\sum_{i\in\mathcal{I}}
\kappa_{i,m} \geq \sum_{i\in \mathcal{I}} R_i +\delta\label{eq:301}
\end{equation}
 for $\delta > 0$ and suppose also that
\begin{align}
& \limsup_{m\rightarrow \infty} I(S_{\mathcal{I},m}; B_me_m(S_{1,m}, \ldots,S_{T,m})|B_m)/m \nonumber\\
&\leq \max\left\{0,\sum_{i\in \mathcal{I}} R_i - (n-\mu)\right\}.
\label{eq:201}
\end{align}
By Eq.\ (\ref{eq:201}) there exists a sequence of $\mu m \times mn$ matrices $b_m$ such that
\begin{align}
& \limsup_{m\rightarrow \infty} I(S_{\mathcal{I},m}; b_me_m(S_{1,m}, \ldots,S_{T,m}))/m \nonumber \\
&\leq\limsup_{m\rightarrow \infty} I(S_{\mathcal{I},m}; B_me_m(S_{1,m}, \ldots,S_{T,m})|B_m)/m \nonumber\\ &\leq \max\left\{0,\sum_{i\in \mathcal{I}} R_i - (n-\mu)\right\}.
\label{eq:202}
\end{align}
For every $m$, define $a_m$ to be an $\mu m \times \mu m$ matrix and $c_m$ to be an $mn \times mn$ matrix such that $a_m b_m c_m$ is a matrix of
the horizontal concatenation of the $\mu m \times \mu m$
identity matrix and the zero matrix.
By Eq.\ (\ref{eq:202}) and the data processing inequality we have
\begin{align}
& \limsup_{m\rightarrow \infty} I(S_{\mathcal{I},m}; a_mb_mc_me_m(S_{1,m}, \ldots,S_{T,m}))/m \nonumber\\
&\leq \max\left\{0,\sum_{i\in \mathcal{I}} R_i - (n-\mu)\right\}.
\label{eq:203}
\end{align}
Define $X_m^{(1)}$ as the first $\mu m$ components in the random vector $e_m(S_{1,m}$, \ldots,
$S_{T,m})$,
and $X_m^{(2)}$ as the remaining components in $e_m(S_{1,m}$, \ldots,
$S_{T,m})$.
We have
\begin{align*}
&H(S_{\mathcal{I},m}|e_m(S_{1,m}, \ldots, S_{T,m}))
\\
&= H(S_{\mathcal{I},m}|X_m^{(1)}, X_m^{(2)})\\
&= H(S_{\mathcal{I},m}) - I(S_{\mathcal{I},m}; X_m^{(1)}, X_m^{(2)})\\
&= H(S_{\mathcal{I},m}) - I(S_{\mathcal{I},m};X_m^{(1)}) - I(S_{\mathcal{I},m};X_m^{(2)}|X_m^{(1)})\textrm{ (by the chain rule)} \\
&= H(S_{\mathcal{I},m}) - I(S_{\mathcal{I},m}; a_m b_m c_me_m(S_{1,m},\ldots,S_{T,m}))\\
&\qquad  - I(S_{\mathcal{I},m};X_m^{(2)}|X_m^{(1)})\\
&\geq H(S_{\mathcal{I},m}) - I(S_{\mathcal{I},m}; b_m e_m(S_{1,m},\ldots,S_{T,m})) - I(S_{\mathcal{I},m};X_m^{(2)}|X_m^{(1)})\\
&\geq  H(S_{\mathcal{I},m}) - I(S_{\mathcal{I},m}; b_me_m(S_{1,m},\ldots,S_{T,m})) - H(X_m^{(2)})\\
&\geq  m\sum_{i\in\mathcal{I}} \kappa_{i,m} \log q - I(S_{\mathcal{I},m}; b_me_m(S_{1,m},\ldots,S_{T,m})) \\
&\qquad - m(n-\mu)\log q\\
&\geq m\Bigl[\Bigl(\sum_{i\in\mathcal{I}} \kappa_{i,m} - n+\mu\Bigr) \log q  - I(S_{\mathcal{I},m}; b_me_m(S_{1,m},\ldots,S_{T,m}))/m\Bigr]\\
&\geq m\Bigl[\Bigl(\sum_{i\in\mathcal{I}} \kappa_{i,m} - n+\mu\Bigr) \log q\\
&\mbox{ }  - I(S_{\mathcal{I},m}; B_me_m(S_{1,m},\ldots,S_{T,m})|B_m)/m\Bigr],
\end{align*}
where $H$ denotes the (conditional) entropy as defined in \cite{cover06}.
By abuse of notation, re-define $\alpha_{\mathcal{I}}$ to
be the projection from
$\prod_{i=1}^{T} \mathbf{F}_q^{k_i}$ to $\prod_{i\in\mathcal{I}} \mathbf{F}_q^{k_i}$
for $\emptyset \neq \mathcal{I} \subseteq \{1$, \ldots, $T\}$.
By using Fano's inequality \cite[Theorem 2.10.1]{cover06}
\begin{align}
& \mathrm{Pr}[(S_{1,m},\ldots, S_{T,m}) \neq d_m(e_m(S_{1,m},\ldots,S_{T,m}))]\nonumber\\
&\geq\mathrm{Pr}[S_{\mathcal{I},m} \neq \alpha_{\mathcal{I}}(d_m(e_m(S_{1,m},\ldots,S_{T,m})))]\nonumber\\
&\geq\frac{H(S_{\mathcal{I},m}|e_m(S_{1,m},\ldots,S_{T,m}))-1}{\log |\prod_{i\in\mathcal{I}}\mathbf{F}_q^{m\kappa_{i,m} }|}\nonumber\\
&\geq \frac{1}{\sum_{i\in\mathcal{I}} \kappa_{i,m}\log q}
\Biggl[\Bigl(\sum_{i\in\mathcal{I}} \kappa_{i,m} - n+\mu\Bigr) \log q \nonumber\\
&\qquad - \frac{1}{m}-\frac{I(S_{\mathcal{I},m}; B_me_m(S_{1,m},\ldots,S_{T,m})|B_m)}{m}\Biggr]. \label{eq401}
\end{align}
By Eqs.\ (\ref{eq:301}), (\ref{eq:201}) and (\ref{eq401}) we can see that $\limsup_{m\rightarrow\infty} \mathrm{Pr}[(S_{1,m}$,
\ldots, $S_{T,m}) $ $\neq$  $d_m(e_m(S_{1,m}$, \ldots, $S_{T,m}))]$ 
$\geq$ $ \delta/(\delta+\sum_{i\in\mathcal{I}} R_i) $ $>$ $0$.
This shows that the limit of mutual information
$I(S_{\mathcal{I},m};$ $B_me_m(S_{1,m}$, \ldots, $S_{T,m})|B_m)/n$
cannot be lower than $\sum_{i\in\mathcal{I}} R_i - (n-\mu)$
while keeping the sum of information rates
$\sum_{i\in\mathcal{I}} \kappa_i$ strictly larger than
$\sum_{i\in\mathcal{I}} R_i$.
\qed

\section{Conclusion}\label{sec4}
In the secure network coding,
there was loss of information rate due to inclusion of random bits at the source node.
In this paper, we have shown that a method to eliminate that loss of information rate
by using multiple statistically independent messages to be kept secret from an eavesdropper,
and called the proposed scheme \emph{secure multiplex network coding.}
The proposed scheme is an adaptation of Yamamoto et~al.'s secure multiplex coding \cite{yamamoto05}
to the secure network coding \cite{cai02b,caiyeung11,silva08}.

\section*{Acknowledgment}
The authors thank anonymous reviewers of NetCod 2011 for
carefully reading the initial manuscript and pointing out
its shortcomings.
The first author would like to thank Prof.\ Hirosuke Yamamoto to teach him
the secure multiplex coding,
Dr.\ Shun Watanabe to point out the relation between
the proposed scheme and \cite{harada08},
Mr.\ Jun Kurihara to point out the relation between
the proposed scheme and \cite{silva09},
Dr.\ Jun Muramatsu and Prof.\ Tomohiro Ogawa for the helpful discussion on the universal
coding.
A part of this research was done during the first author's stay
at the Institute of Network Coding, the Chinese University
of Hong Kong,
and he greatly appreciates the hospitality by Prof.\ Raymond
Yeung.
This research was partially supported by 
the MEXT Grant-in-Aid for Young Scientists (A) No.\ 20686026 and
(B) No.\ 22760267, and Grant-in-Aid for Scientific Research (A) No.\ 23246071.
The Center for Quantum Technologies is funded
by the Singapore Ministry of Education and the National Research
Foundation as part of the Research Centres of Excellence programme.

\appendices
\section{Proof of Theorem \ref{thm2}}\label{app:a}
In order to show Theorem \ref{thm2},
we introduce the following lemma.
\begin{lemma}\label{lem1}
Under the same assumption as Theorem \ref{thm2},
we have
\begin{equation}
\mathbf{E}_f \exp(-\rho H(F(X);Z|F=f))  \leq   
|\mathcal{S}|^{-\rho} + \mathbf{E}[P_{X|Z}(X|Z)^\rho]\label{eq:lem1}
\end{equation}
for $0\leq \rho \leq 1$.
\end{lemma}

\noindent\emph{Proof of Theorem \ref{thm2}.}
\begin{align*}
&  \mathbf{E}_f \exp(\rho I(F(X);Z|F=f))\\
&= \mathbf{E}_f \exp(\rho H(F(X)|F=f) - \rho H(F(X);Z|F=f))\\
&\leq \mathbf{E}_f |\mathcal{S}|^\rho \exp(-\rho H(F(X);Z|F=f))\\
&\leq \mathbf{E}_f |\mathcal{S}|^\rho (|\mathcal{S}|^{-\rho} + \mathbf{E}[P_{X|Z}(X|Z)^\rho]) \textrm{ (by Eq. (\ref{eq:lem1}))}\\
&= 1+ |\mathcal{S}|^\rho\mathbf{E}[P_{X|Z}(X|Z)^\rho] \tag*{\qed}
\end{align*}

\noindent\emph{Proof of Lemma \ref{lem1}.}
Fix $z \in \mathcal{Z}$.
The concavity of $x^\rho$ for $0\leq\rho\leq 1$ implies 
\begin{align}
& \mathbf{E}_f \sum_{s \in \mathcal{S}} P_{f(X)|Z}(s|z)^{1+\rho} \nonumber\\
&= \sum_{x\in \mathcal{X}} P_{X|Z}(x|z) \mathbf{E}_f \Bigl( \sum_{x'\in f^{-1}(x)} P_{X|Z}(x'|z)\Bigr)^\rho\nonumber\\
&\leq  \sum_{x\in \mathcal{X}} P_{X|Z}(x|z) \Bigl(\underbrace{\mathbf{E}_f  \sum_{x'\in f^{-1}(x)} P_{X|Z}(x'|z)}_{(*)}\Bigr)^\rho. \label{eq200}
\end{align}
Since $f$ is chosen from a family of two-universal hash functions
defined in Definition \ref{def:twouniv}, we have
\begin{align*}
(*) &\leq P_{X|Z}(x|z) + \sum_{x \neq x' \in \mathcal{X}} \frac{P_{X|Z}(x'|z)}{|\mathcal{S}|}\\
&\leq P_{X|Z}(x|z) + |\mathcal{S}|^{-1}.
\end{align*}
Since any two positive numbers $x$ and $y$ satisfy
$(x+y)^\rho \leq x^\rho + y^\rho$ for $0 \leq \rho \leq 1$,
we have
\begin{equation}
(P_{X|Z}(x|z) + |\mathcal{S}|^{-1})^\rho \leq P_{X|Z}(x|z)^\rho + |\mathcal{S}|^{-\rho}.
\label{eq201}
\end{equation}
By Eqs.\ (\ref{eq200}) and (\ref{eq201}) we can see
\[
\mathbf{E}_f \sum_{s \in \mathcal{S}} P_{f(X)|Z}(s|z)^{1+\rho} \leq
\sum_{x \in \mathcal{X}} P_{X|Z}(x|z)^{1+\rho} + |\mathcal{S}|^{-\rho}.
\]
Taking the average over $Z$ of the both sides
of the last equation, we have
\begin{equation}
\mathbf{E}_f \mathbf{E}_{XZ} P_{f(X)|Z}(f(X)|Z)^{\rho} \leq
\mathbf{E}_{XZ}  P_{X|Z}(X|Z)^{\rho} + |\mathcal{S}|^{-\rho}. \label{eq203}
\end{equation}
Define $g(\rho) = \mathbf{E}_{XZ} P_{f(X)|Z}(f(X)|Z)^{\rho}$
as a function of $\rho$ with fixed $f$ and $P_{XZ}$, and
$h(\rho) = \log g(\rho)$.
We have
\begin{align*}
g'(\rho)
&= \mathbf{E}_{XZ} P_{f(X)|Z}(f(X)|Z)^{\rho} \log P_{f(X)|Z}(f(X)|Z),\\
g''(\rho)
&= \mathbf{E}_{XZ} P_{f(X)|Z}(f(X)|Z)^{\rho} (\log P_{f(X)|Z}(f(X)|Z))^2,\\
h'(\rho) &= g'(\rho)/g(\rho),\\
h''(\rho) &= \frac{g''(\rho)g(\rho) - [g'(\rho)]^2}{g(\rho)^2}.
\end{align*}
Define $(X'$, $Z')$ to be the random variables that have the same
joint distribution as $(X,Z)$ and statistically independent of
$X$ and $Z$.
To examine the sign of $h''(\rho)$ we compute 
\begin{align*}
&  g''(\rho)g(\rho) - [g'(\rho)]^2\\
&= \mathbf{E}_{XZX'Z'} P_{f(X)Z}(f(X),Z)^{\rho}P_{f(X)Z}(f(X'),Z')^{\rho}  \\*
&\qquad [(\log P_{f(X)|Z}(f(X)|Z))^2  - \log P_{f(X)|Z}(X|Z)\log P_{f(X)|Z}(X'|Z')]\\
&= \frac{1}{2} \mathbf{E}_{XZX'Z'} P_{f(X)Z}(f(X),Z)^{\rho}P_{f(X)Z}(f(X'),Z')^{\rho}
\\*
&\qquad[(\log P_{f(X)|Z}(f(X)|Z))^2
+ (\log P_{f(X)|Z}(f(X')|Z'))^2 \\*
&\qquad - 2\log P_{f(X)|Z}(f(X)|Z)\log P_{f(X)|Z}(f(X')|Z')]\\
&= \frac{1}{2} \mathbf{E}_{XZX'Z'} P_{f(X)Z}(f(X),Z)^{\rho}P_{f(X)Z}(f(X'),Z')^{\rho}\\*
&\qquad [\log P_{f(X)|Z}(f(X)|Z) - \log P_{f(X)|Z}(f(X')|Z')]^2\\
&\geq 0.
\end{align*}
This means that $h''(\rho) \geq 0$ and $h(\rho)$ is convex.
We can see 
\begin{align}
\mathbf{E}_{XZ} P_{f(X)|Z}(f(X)|Z)^{\rho} &= \exp(h(\rho)) \nonumber\\
&\geq  \exp(\underbrace{h(0)}_{=0} + \rho h'(0))\nonumber\\
&= \exp(-\rho H(f(X)|Z)). \label{eq204}
\end{align}
By Eqs.\ (\ref{eq203}) and (\ref{eq204}) we see that
Eq.\ (\ref{eq:lem1}) holds. \qed



\section{Notes added after publication}\label{app:b}
This appendix is a note after submission of the final version to
Proc.\ NetCod 2011.
In Section \ref{sec32} we claimed that the mutual information to
Eve can be arbitrary small. This means that the mutual information
can be made exactly zero for every eavesdropping matrix $b$.
The reason is as follows:
For fixed $b$ and $\ell$, we have
\begin{equation}
I(S_{\mathcal{I}};BX|B=b,L=\ell)= H(S_{\mathcal{I}}|B=b,L=\ell) - H (S_{\mathcal{I}}|bX,L=\ell).
\label{eq:mut}
\end{equation}
The first term $H(S_{\mathcal{I}}|B=b,L=\ell)$ is an integer multiple of
$\log q$ since $S_{\mathcal{I}}$ is assumed to have the uniform
distribution. 
For fixed $b$ and $\ell$, we have
$bX = b\ell^{-1}(S_1$, \ldots, $S_{T+1})$.
For a given realization $bx$ of $bX$,
the set of solutions $s$ such that $bx = b\ell^{-1}s$ is written
as $\ker(b\ell^{-1}) + $ some vector $v$.
This means that the set of
 possible candidates of $S_{\mathcal{I}}$ given
realization $bx$ of $bX$ is written as $\alpha_{\mathcal{I}}(\ker(b\ell^{-1})) +
\alpha_{\mathcal{I}}(v)$, and $S_{\mathcal{I}}$ given
realization $bx$ is uniformly distributed on 
$\alpha_{\mathcal{I}}(\ker(b\ell^{-1})) +
\alpha_{\mathcal{I}}(v)$.
Since the cardinality of $\alpha_{\mathcal{I}}(\ker(b\ell^{-1})) +
\alpha_{\mathcal{I}}(v)$ is independent of $X$ for fixed $b$ and $\ell$,
the second term $H (S_{\mathcal{I}}|bX,L=\ell)$ is also an
integer multiple of $\log q$. Therefore, if Eq.\ (\ref{eq:ub8})
holds for every $B$ as verified in Section \ref{sec32} and
the RHS of Eq.\ (\ref{eq:ub8}) is $<\log q$, then
the LHS of Eq.\ (\ref{eq:ub8}) must be zero.

In Section \ref{sec1} we overlooked the relevant research
result by Cai (``Valuable messages and random outputs of channels in linear network coding,'' Proc.\ ISIT 2009, Seoul, Korea, Jun. 2009, pp. 413--417).
Cai proved that random linear network coding gives the strongly
secure network coding in the sense of \cite{harada08} with
arbitrarily high probability with sufficiently large finite
fields. The advantages of the present result over Cai's result are
(1) we do not have to change encoding at intermediate nodes,
(2) our construction is universal secure in the sense of \cite{silva09,silva08},
and (3) much smaller finite fields can be used than Cai's result.
\end{document}